\documentstyle[12pt]{article}
\def\theequation{\arabic{equation}}

\renewcommand{\thefootnote}{\fnsymbol{footnote}}
\begin{document}
\thispagestyle{empty}

\begin{center}
{\large{\bf SUPERWEYL COCYCLE IN d=4
 AND SUPERCONFORMAL-INVARIANT OPERATOR}} \vspace{1.5cm}\\

{\large R.P.Manvelyan\footnote{E-MAIL:manvelyan@vxc.yerphi.am
}}\vspace{0.5cm}\\

{\it Theoretical Physics Department,}\\
{\it Yerevan Physics Institute}\\
{\it Alikhanyan Br. st.2, Yerevan, 375036 Armenia }
\end{center}
\bigskip
\bigskip
\bigskip
\bigskip
\begin{abstract}
The super-Weyl cocycle (effective action for supertrace anomaly)
and corresponding invariant operator in nonminimal
formulation of $d=4$,$N=1$  supergravity are obtained.
\end{abstract}
\vfill
\setcounter{page}0
\renewcommand{\thefootnote}{\arabic{footnote}}
\setcounter{footnote}0
\renewcommand{\theequation}{\arabic{section}.\arabic{equation}}
\setcounter{equation}{0}
\newpage
\newcommand{\ra}{\rightarrow}

\pagestyle{plain}

\section{\bf Introduction.}
\setcounter{equation}{0}

In previous papers \cite{KMM} it was shown that the Weyl cocycle in $d=4,6$
has a form of Liouville action in $d=2$ \cite{Pol}:
\begin{equation}\label{1}
S = \int d^dx\sqrt{g}[\sigma\Delta_{d}\sigma +
\sigma\times{\it Anomaly}]
\end{equation}
where $\Delta_{d}$ is the zero weight conformal-invariant operator of $d$-th
order on derivatives.
This is true also for $d=2$ supersymmetric Weyl cocycle \cite{Dav}
(Super-Liouville action).
In this letter we consider the supersymmetric extension of Weyl cocycle in
$d=4$, using the set of the superspace constraints,leading to
non-minimal formulation of $d=4, N=1$ supergravity \cite{LJ}.
The different formulations of superfield supergravity  in $d=4$ differ by their
auxiliary-field structure and are parametrized by real number parameter $n$
in the generalized superconformal constraint\cite{LJ} \cite{Gates}
(we use the notation of ref. \cite{Sieg}).
\begin{eqnarray}
\label{2}
T_{\alpha \beta}^{\quad C} & = & T_{\alpha
\dot{\beta}}^{\quad C} +
2\imath\sigma^c_{\alpha\dot{\beta}} = T_{a b}^{\quad c} = T_{\alpha a}^{\quad
b}
 - \frac{1}{4}\sigma_{a \alpha \dot{\beta}}\sigma^{b \gamma
\dot{\beta}}T_\gamma = 0;\nonumber \\
  n \neq  & -\frac{1}{3}&: R - \frac{n+1}{3n+1}
\nabla_{\dot{\alpha}}T^{\dot{\alpha}} + \left(\frac{n+1}{3n+1}\right)^2
T_{\dot{\alpha}}T^{\dot{\alpha}} = 0 ,\\ n  = & -\frac{1}{3}&: T_\alpha = 0
\nonumber
\end{eqnarray}
where
\begin{equation}
\label{2a}
 T_{\alpha} = T_{\alpha a}^{\quad a} ,\quad
 R = \frac{1}{12}\sigma^{\gamma \dot{\alpha}}_a\sigma_{b
\gamma}^{\quad\dot{\beta}}
 R_{\dot{\alpha}\dot{\beta}}^{\quad a b}
\end{equation}

If $n=-1/3$ we have the minimal Wess-Zumino \cite{WZ} formulation
of supergravity with Howe and Tucker's \cite{HT} superconformal group,
\begin{eqnarray}
\label{3}
 E_\alpha^{\,\,M} &\ra & e^{\frac{1}{2}(2\tilde\Sigma - \Sigma)}
 E_\alpha^{\,\,M} \\
\nabla_{\dot{\alpha}}\Sigma = 0  &,& \nabla_{\alpha}\tilde\Sigma = 0 \nonumber
\end{eqnarray}
with chiral conformal parameter $\Sigma$.
The effective action for this formulation was constructed in 1988 by
I.L.Buchbinder and S.M.Kuzenko \cite{BK} ,but this effective action is
not of the form:
\begin{equation}
\label{4}
\int {\it Anomaly}\frac{1}{\Delta_d}{\it Anomaly}
\end{equation}
and corresponding cocycle is not second order on super-Weyl parameter.
Therefore in this formulation there is no room for the supersymmetric
 extension of our construction for conformal invariant operator.
 So in our paper we use $n=-1$ formulation of supergravity constraints with
 superconformal transformation including linear scalar superfield parameter.
 \cite{LJ,Gates}.
\begin{eqnarray}
\label{5}
 E_\alpha^{\,\,M} &\ra & e^{L}E_\alpha^{\,\,M} \\
\nabla_{\dot{\alpha}}\nabla^{\dot{\alpha}}L = 0  &,&
\nabla_{\alpha}\nabla^{\alpha}L = 0 \nonumber
\end{eqnarray}
In this formulation we can write down quadratic supercocycle and
superconformal invariant fourth-order differential operator,which is the
supersymmetric extension of corresponding Weyl invariant operator obtained
 in the work of Riegert \cite{Rieg}.
\begin{equation}
\label{6}
\sqrt{g}\Delta_4 =\sqrt{g}( \hat{{\Box}}^2 -
2\hat{R}^{\mu\nu}\hat{\nabla}_{\mu}
\hat{\nabla}_{\nu}
+ \frac{2}{3}\hat{R}\hat{\Box} - \frac{1}{3}(\hat{\nabla}^{\mu} \hat{R})
\hat{\nabla}_{\mu})
\end{equation}
where $\hat{\nabla}$ and $\hat{R}$ are ordinary covariant derivative
and curvature.

The 1-supercocycle of super-Weyl group is the super-Weyl variation of
corresponding anomalous effective action for superconformal matter superfield
in external supergravity:
\begin{equation}
\label{7}
{\bf S}(L,E_{\alpha}^{\,\, M}) = W(e^{L} E_{\alpha}^{\,\, M})
- W(E_{\alpha}^{\,\, M})
\end{equation}
and has to satisfy the cocyclic property:
\begin{equation}
\label{8}
{\bf S}(L_1 + L_2;E_{\alpha}^{\,\, M}) =
{\bf S}(L_1; e^{L_2}E_{\alpha}^{\,\, M})
+ {\bf S}(L_2; E_{\alpha}^{\,\, M})
\end{equation}
Finally, let's note that from (\ref{8}) the term of the highest-
order over group parameter $L$ in the cocycle has to be super-Weyl invariant
\cite{KMM}.
Therefore the quadratic over $L$ invariant action  containing
(super)conformal-invariant operator $\Delta$ leads to the following cocycle:
\begin{equation}
\label{8a}
{\bf S} \sim \int \left(\frac{1}{2}L\Delta L - A(E)L\right)
\end{equation}
where $A(E)$ is the possible contribution to the super-Weyl anomaly \cite{KMM}
satisfying the well-known Wess-Zumino consistency condition \cite{WZ1},which
is the algebraic version of cocyclic property .
We can obtain the linear part $A(E)L$ solving the equation which follows
from (\ref{8}) and (\ref{8a})
\begin{equation}
\label{8b}
 \Omega\Delta L = \delta_{\Omega}A(E)L
\end{equation}
Here $\delta_{\Omega}$ is super-Weyl transformation with infinitesimal
linear superfield parameter $\Omega$.

\section{\bf Supercocycle and superconformal-invariant operator}
\setcounter{equation}{0}

Firstly let's remind that in the $n=-1$ formulation of supergravity
constraints we can express all supertorsions and supercurvatures in terms of
following superfields (due to (\ref{2}) , and Bianchi identities):
\begin{equation}
\label{9}
  T_{\alpha},\quad  G_{a}=\frac{1}{12}\sigma^{\alpha\dot{\beta}}_{a}
  (R^\gamma_{\,\,\dot{\beta}\alpha\gamma} +
  R^{\dot\gamma}_{\,\,\alpha\dot{\beta}\dot{\gamma}}),\quad
  W_{\alpha\beta\gamma} = \sigma^{a}_{(\alpha\dot{\delta}}
  \sigma^{b\dot{\delta}}_{\beta}T_{ab\gamma)}
\end{equation}
where
\begin{eqnarray}
\label{10}
& & \nabla_{(\alpha}T_{\beta} = 0,\quad G_a = \tilde{G}_a,\quad
 \left(\nabla_{\dot{\alpha}} - \frac{1}{2}\tilde{T}_{\dot{\alpha}}
 \right)W_{\alpha\beta\gamma},\nonumber\\
& & \sigma^a_{\alpha\dot{\alpha}}T_{aD}^{\quad D}
 = \frac{\imath}{4}\left(\nabla_{\dot{\alpha}}T_\alpha +
 \nabla_\alpha T_{\dot{\alpha}}\right),\quad T_{\alpha B}^{\quad B} = T_\alpha
\end{eqnarray}
This set of superfields has the following local superscale
transformations \cite{LJ}:
\begin{eqnarray}
\label{11}
\delta_{\Omega}W_{\alpha\beta\gamma}&=& 3\Omega W_{\alpha\beta\gamma},
\quad \delta_{\Omega}T_\alpha = \Omega T_\alpha + 6\nabla_\alpha \Omega \\
\delta_{\Omega}G_a &=& 2\Omega G_a + \frac{1}{4}\sigma^{\alpha\dot{\beta}}_a
\left[\nabla_{\dot{\beta}},\nabla_{\alpha}\right]\Omega \nonumber
\end{eqnarray}
Beside that we can derive the infinitesimal form of transformations
of the supercovariant derivatives acting on the
zero weight scalar linear superfield \cite{LJ,Gates}:
\begin{eqnarray}
\label{12}
\delta_{\Omega}\nabla_{\alpha} &=& \Omega\nabla_{\alpha},\quad
\delta_{\Omega}\nabla_{\dot{\alpha}} = \Omega\nabla_{\dot{\alpha}}\nonumber \\
\delta_{\Omega}\nabla_{\alpha\dot{\alpha}} &=&
2\Omega\nabla_{\alpha\dot{\alpha}} -
\frac{\imath}{2}\left(\nabla_{\alpha}\Omega \nabla_{\dot{\alpha}} +
\nabla_{\dot{\alpha}}\Omega \nabla_{\alpha}\right)
\end{eqnarray}
where $\nabla_{\alpha\dot{\alpha}} = \sigma^{a}_{\alpha\dot{\alpha}}\nabla_{a}$

{}From (\ref{11}) and (\ref{12}) it follows that the supertorsion component
  $T_\alpha$ plays  a role of the
gauge field for the super-Weyl transformation.The  presence of this phenomenon
in $n=-1$ formulation leads us to a very clear construction of
superconformal-invariant operator .In the minimal $(n=-1/3)$ formulation
(and in nonsupersymmetric case too) there is not such a gauge field.
 Owing to this it is easy to define from (\ref{11}) and (\ref{12})
 super-Weyl covariant superderivative  ${\bf D}_{\alpha \dot{\beta}}$:
 \begin{eqnarray}
\label{13}
{\bf D}_{\alpha\dot{\alpha}} &=& \nabla_{\alpha\dot{\alpha}} -
\frac{\imath}{12}\left(T_{\alpha}\nabla_{\dot{\alpha}} +
T_{\dot{\alpha}}\nabla_{\alpha}\right)\\
\delta_{\Omega} {\bf D}_{\alpha\dot{\alpha}} &=&
2\Omega{\bf D}_{\alpha\dot{\alpha}}\nonumber
\end{eqnarray}
acting on a scalar linear superfield of zero superconformal weight.
Two other supercovariant derivatives $ \nabla_\alpha , \nabla_{\dot{\alpha}}$
are automatically superscale covariant on the zero weight scalar.
Then we can conclude that the following expression:
\begin{equation}
\label{14}
\frac{1}{2}E^{-1}{\bf D}_{\alpha\dot{\alpha}}L{\bf D}^{\alpha\dot{\alpha}}L
\end{equation}
is superconformal invariant
\begin{equation}
\label{14a}
\delta_{\Omega}\left[\frac{1}{2}E^{-1}{\bf D}_{\alpha\dot{\alpha}}L
{\bf D}^{\alpha\dot{\alpha}}L\right] = 0
\end{equation}
and has to be the highest term of our cocycle.
The linear on $L$ term we can construct from the cocyclic property (\ref{8b})
and from the following formula:
\begin{equation}
\label{15}
\delta_{\Omega}\left[E^{-1}\frac{\imath}{12}
\left(\nabla_{\alpha}T_{\dot{\alpha}}
 + \nabla_{\dot{\alpha}}T_{\alpha}\right){\bf D}^{\alpha\dot{\alpha}}L\right] =
 E^{-1}{\bf D}_{\alpha\dot{\alpha}}\Omega{\bf D}^{\alpha\dot{\alpha}}L
\end{equation}
Then we can write out super-Weyl cocycle comparing (\ref{15}) and (\ref{14a})
to (\ref{8a}),(\ref{8b}):
\begin{equation}
\label{16}
 {\bf S} = \int d^8zE^{-1}\left[\frac{1}{2}{\bf D}_{\alpha\dot{\alpha}}L
{\bf D}^{\alpha\dot{\alpha}}L -
\frac{\imath}{12}\left(\nabla_{\alpha}T_{\dot{\alpha}}
 + \nabla_{\dot{\alpha}}T_{\alpha}\right){\bf D}^{\alpha\dot{\alpha}}L\right]
\end{equation}

After that we can obtain  the expressions for $\Delta_{4}^{SUSY}$ and
$A_{4}^{SUSY}$ after integration by parts in (\ref{16}) using the following
relation:
\begin{equation}
\label{17}
\int d^8zE^{-1} \nabla_{C}V^{C}(-1)^C = \int d^8zE^{-1}V^{C}
T_{CD}^{\quad D}(-1)^{D} \neq 0
\end{equation}
in $n=-1$ formulation due to (\ref{10}).
It leads us to the following final results:
\begin{equation}
\label{18}
\Delta_{4}^{SUSY} = {\bf D}_{\alpha\dot{\alpha}}{\bf D}^{\alpha\dot{\alpha}} +
\frac{\imath}{3}\left(\nabla_{\alpha}T_{\dot{\alpha}}
 + \nabla_{\dot{\alpha}}T_{\alpha}\right){\bf D}^{\alpha\dot{\alpha}}
\end{equation}
\begin{eqnarray}
\label{19}
A_{4}^{SUSY} &=& \frac{1}{36}\left(\nabla_{\alpha}T_{\dot{\alpha}}
 + \nabla_{\dot{\alpha}}T_{\alpha}\right)\left(\nabla^{\alpha}T^{\dot{\alpha}}
 + \nabla^{\dot{\alpha}}T^{\alpha}\right)\\ &-& \frac{\imath}{12}
 {\bf D}_{\alpha\dot{\alpha}}\left(\nabla^{\alpha}T^{\dot{\alpha}}
 + \nabla^{\dot{\alpha}}T^{\alpha}\right)\nonumber
\end{eqnarray}
This operator is actually fourth-order (on the component level),over space-time
derivatives, because the linear scalar multiplet $L$
includes the second derivatives of its first component.

In conclusion let's note that owing to (\ref{17}) the expression (\ref{19})
obtaining after integration by parts is not full derivative and has to include
non-trivial contribution in anomaly (super-Euler characteristic density).

{\bf{Acknowledgments}}
The author thanks R.L.Mkrtchyan and D.R.Kara\-kha\-nyan for useful discussions.
This work was supported in part by the INTAS grant \# 93-1038 and by
the  grant YPI-1995 of the "Bundesminister f\"{u}r Forschung und Technologie",
Federal Republic of Germany.


\end{document}